\documentclass[12pt]{elsart}
\usepackage{feynmp,epsfig,graphics}

\def\tr{{\rm tr} \,}

\def\pslash{\FMslash p}

\begin{document}
GSI-Preprint-2003-18
\begin{frontmatter}
\title{Self consistent and covariant propagation \\ of pions, nucleon and isobar resonances \\
in cold nuclear matter }
\author[PTE]{C.L.\ Korpa}
\author[GSI,TU]{and M.F.M. Lutz}
\address[PTE]{Department of Theoretical Physics, University of
Pecs, \\Ifjusag u.\ 6, 7624 Pecs, Hungary}
\address[GSI]{Gesellschaft f\"ur Schwerionenforschung (GSI),\\
Planck Str. 1, 64291 Darmstadt, Germany}
\address[TU]{Institut f\"ur Kernphysik, TU Darmstadt\\
D-64289 Darmstadt, Germany}
\begin{abstract}
We evaluate the in-medium spectral functions for pions, nucleon
and isobar resonances in a self consistent and covariant manner.
The calculations are based on a recently developed formulation
which leads to predictions in terms of the pion-nucleon scattering
phase shifts and a set of Migdal parameters describing important
short range correlation effects. We do not observe significant
softening of pion modes if we insist on reasonable isobar
resonance properties but predict a considerable broadening of the
$N(1440)$ and $N(1520)$ resonances in nuclear matter. Contrasted
results are obtained for the s-wave $N(1535)$ and $N(1650)$
resonances which are affected by a nuclear environment very
little. The properties of slowly moving isobar's in nuclear matter
are found to depend very sensitively on a soft form factor in the
$\pi NN$ vertex, which is not controlled by the $\pi N$ scattering
data.
\end{abstract}
\end{frontmatter}


\section{Introduction}

The study of pion propagation in dense nuclear matter is of
central importance when addressing the in-medium modifications of
nucleon and delta resonances
\cite{Oset:Weise,Dickhoff,Oset:Salcedo,Thies,Ericson:Weise,Migdal,Nieves:Oset:Recio,Xia:Siemens:Soyeur,Korpa:Malfliet,Knoll,Nakano,Polls,Schramm,Lutz:Migdal}.
There are strong hints from the empirical photon nucleus
absorption cross section as well as from pion-nucleus scattering
data that the low-lying nucleon and delta resonances do change
their properties in nuclear matter substantially already at
nuclear saturation density
\cite{Koch,photo-absorption,Kondratyuk,Alberico,Chen,Boffi}.
Recent data on electroproduction of isobars off helium three
\cite{Richter} are interpreted in terms of a repulsive in-medium
mass change and an increased decay width of the isobar-resonance
\cite{Richter}. Since most nucleon and isobar resonances have a
substantial decay fraction into one nucleon and one pion, the
in-medium resonance structure reflects to a large extent the
medium modified propagation properties of pions. The pion self
energy in nuclear matter is quantitatively constrained  by pionic
atom data
\cite{Migdal,Batty:Fridmann:Gal,Kolomeitsev:Kaiser:Weise}. Most
exciting are the recently established states where a negatively
charged pion is bound by a heavy nucleus in an s-wave or p-wave
state \cite{GSI-pi-atom}. It is still an open problem to find a
quantitative and microscopic derivation in particular for the
large absorptive part in the nuclear optical potential needed in
the phenomenological description of pionic atom data
\cite{Migdal,Batty:Fridmann:Gal,Kolomeitsev:Kaiser:Weise}. The
problem requires a non-perturbative many-body approach, since the
low-density expansion ceases to converge rapidly at the relevant
nuclear densities even when chiral correction terms are considered
\cite{Kaiser:Weise,Park:Jung:Min,Meissner:Oller:Wirzba}.

In this work we generalize the covariant framework of \cite{LuKo},
which was recently proposed for the self consistent propagation of
antikaons and hyperon-resonances in nuclear matter, to the problem
of pion, nucleon- and isobar-resonance propagation incorporating
short range correlation effects. The merit of our scheme is that
it is formulated entirely in terms of the two-body scattering
phase-shifts properly extrapolated to subthreshold energies. Given
the empirical $\pi N$ phase shifts and values for the Migdal
parameters \cite{Migdal,Nakano,Polls,Schramm} describing important
short range correlations the scheme is parameter free. We expect
self consistency to lead to a broadening of the pion spectral
function which may help to establish a microscopic understanding
of pionic atom data and also offer an improved understanding of
the expected broadening of nucleon and isobar resonances in
nuclear matter. Our present approach constitutes a significant
progress as compared to previous self consistent calculations
\cite{Xia:Siemens:Soyeur,Korpa:Malfliet,Knoll} which were based on
p-wave pion-nucleon dynamics only. Moreover this work is the first
attempt to consider the effects of the in-medium mixing of partial
wave amplitudes. In particular the feedback effect of the
in-medium modified s-wave $N(1520)$ and $N(1650)$, the p-wave
$N(1440)$ and the d-wave $N(1535)$ resonances to the propagation
properties of pions will be addressed in this work. Similarly the
isobar resonances s-wave $\Delta(1620)$, p-wave $\Delta (1232)$
and $\Delta (1600)$ and the d-wave $\Delta (1700)$ resonances are
considered.

\section{Self consistent nuclear pion dynamics}

We briefly recall the self consistent and covariant many-body framework
introduced in \cite{LuKo} appropriately adjusted to pion propagation in nuclear matter.
First we recall the vacuum on-shell pion-nucleon scattering amplitude
\begin{eqnarray}
&&\langle \pi ^{j}(\bar q)\,N(\bar p)|\,T\,| \pi ^{i}(q)\,N(p) \rangle
=(2\pi)^4\,\delta^4(q+p-\bar q-\bar p)\,
\nonumber\\
&& \qquad \qquad   \qquad \qquad  \qquad \qquad \times \,\bar u(\bar p)\,
T^{ij}_{\pi  N \rightarrow \pi  N}(\bar q,\bar p ; q,p)\,u(p) \,,
\label{on-shell-scattering}
\end{eqnarray}
where  $\delta^4(..)$ guarantees energy-momentum conservation and
$u(p)$ is the nucleon isospin-doublet spinor. The vacuum
scattering amplitude is decomposed into its isospin channels.
Applying standard notation for the Paul matrices $\tau_i$ we write
\begin{eqnarray}
&&T^{i j}_{\pi  N \to \pi  N}(\bar q,\bar p \,; q,p) =
T^{(1/2)}(\bar k,k;w)\,P^{ij}_{(I=1/2)}+ T^{(3/2)}(\bar
k,k;w)\,P^{ij}_{(I=3/2)}\;,
\nonumber\\
&& P^{ij}_{(I=1/2)}= \frac{1}{3}\, \tau^i\, \tau^j \, , \quad
P^{ij}_{(I=3/2)}= \delta^{ij}\,1-\frac{1}{3}\, \tau^i\, \tau^j \;,
\label{}
\end{eqnarray}
where $q, p, \bar q, \bar p$ are the initial and final pion and
nucleon 4-momenta and
\begin{eqnarray}
w = p+q = \bar p+\bar q\,,
\quad k= \half\,(p-q)\,,\quad
\bar k =\half\,(\bar p-\bar q)\,.
\label{def-moment}
\end{eqnarray}
In quantum field theory the scattering amplitudes  follow as the
solution of the Bethe-Salpeter matrix equation
\begin{eqnarray}
T(\bar k ,k ;w ) &=& K(\bar k ,k ;w )
+\int \frac{d^4l}{(2\pi)^4}\,K(\bar k , l;w )\, G(l;w)\,T(l,k;w )\;,
\nonumber\\
G(l;w)&=&-i\,S ({\textstyle {1\over 2}}\,w+l)\,D({\textstyle
{1\over 2}}\,w-l) \,, \label{BS-eq}
\end{eqnarray}
in terms of the Bethe-Salpeter kernel $K(\bar k,k;w)$, the free
space nucleon propagator $S (p)=1/(\pslash-m_N+i\,\epsilon)$ and
pion propagator $D(q)=1/(q^2-m_\pi^2+i\,\epsilon)$. Following
\cite{LK} we neglect self energy corrections in the nucleon and
pion propagators. In a chiral scheme such effects are of
subleading order. The Bethe-Salpeter equation (\ref{BS-eq})
properly implements  Lorentz invariance and unitarity for the
two-body scattering process. The generalization of (\ref{BS-eq})
to a coupled-channel system is straightforward.

The pion-nucleon scattering process is readily generalized from the
vacuum to the nuclear matter case. In compact notation we write
\begin{eqnarray}
&& {\mathcal T} = {\mathcal K} + {\mathcal K} \cdot {\mathcal G} \cdot {\mathcal T}  \;,\quad
{\mathcal T} = {\mathcal T}(\bar k,k; w,u) \;, \quad {\mathcal G} = {\mathcal G} (l;w,u) \,,
\label{hatt}
\end{eqnarray}
where the  in-medium scattering amplitude ${\mathcal T}(\bar k,k;w,u)$ and the two-particle
propagator ${\mathcal G}(l;w,u)$ depend now on the 4-velocity $u_\mu$
characterizing the nuclear matter frame. For nuclear matter moving with a velocity
$\vec v$ one has
\begin{eqnarray}
u_\mu =\left(\frac{1}{\sqrt{1-\vec v\,^2/c^2}},\frac{\vec v/c}{\sqrt{1-\vec v\,^2/c^2}}\right)
\;, \quad u^2 =1\,.
\label{}
\end{eqnarray}
We emphasize that (\ref{hatt}) is properly defined from a Feynman
diagrammatic point of view even in the case where the in-medium
scattering process is no longer well defined due to a broad pion
spectral function. We consider the effect of an in-medium modified
two-particle propagator ${\mathcal G}$
\begin{eqnarray}
&& \Delta S  (p,u) = 2\,\pi\,i\,\Theta \big[p\cdot u \big]\,
\delta(p^2-m_N^2)\,\big( \pslash +m_N \big)\, \Theta
\big(k_F^2+m_N^2-(u\cdot p)^2\big)\,,
\nonumber\\
&&{\mathcal S} (p,u) = S (p)+ \Delta S (p,u)\,, \quad {\mathcal
D}(q,u)=\frac{1}{q^2-m_\pi^2-\Pi (q,u)} \;,
\nonumber\\
&& {\mathcal G}(l;w,u) = -i\,{\mathcal S} ({\textstyle {1\over
2}}\,w+l,u)\,{\mathcal D} ({\textstyle {1\over 2}}\,w-l,u) \;,
\label{hatg}
\end{eqnarray}
where the Fermi momentum $k_F$ parameterizes the nuclear density
$\rho$ with
\begin{eqnarray}
\rho = -2\,\tr \,\gamma_0\,\int \frac{d^4p}{(2\pi)^4}\,i\,\Delta S
(p,u) = \frac{2\,k_F^3}{3\,\pi^2\,\sqrt{1-\vec v\,^2/c^2}}  \;.
\label{rho-u}
\end{eqnarray}
In the rest frame of the bulk with $u_\mu=(1,\vec 0\,)$ one
recovers with (\ref{rho-u}) the standard result $\rho =
2\,k_F^3/(3\,\pi^2)$. In a more complete approach the effect of
nucleonic correlation and binding on $\Delta S$ should be
considered. This is beyond the scope of this work. The pion self
energy $\Pi (q,u)$ is evaluated self consistently in terms of the
in-medium scattering amplitudes
\begin{eqnarray}
\Pi (q,u) &=&  2\,\tr \int \frac{d^4p}{(2\pi)^4}\,i\,\Delta S
(p,u)\,   \bar {\mathcal T} \big({\textstyle{1\over
2}}\,(p-q), {\textstyle{1\over 2}}\,(p-q);p+q,u \big) \nonumber\\
&+& \Delta \Pi  (q,u) \,, \qquad \bar {\mathcal T} =
\frac{1}{3}\,\hat {\mathcal T}^{(I=1/2)}+ \frac{2}{3}\,\hat
{\mathcal T}^{(I=3/2)} \;, \label{k-self}
\end{eqnarray}
where the in-medium amplitudes, $\hat {\mathcal T}^{(I)}(\bar
k,k;w,u)$, are defined with respect to the free-space interaction
kernel, $K$,
\begin{eqnarray}
&&\hat {\mathcal T}= K+K\cdot {\mathcal G}\cdot \hat {\mathcal T}
= T+T\cdot \Delta {\mathcal G} \cdot \hat {\mathcal T}\;,\quad
\Delta {\mathcal G}={\mathcal G}-G\;. \label{rewrite}
\end{eqnarray}
Additional contributions induced by the in-medium modification of
the interaction kernel ${\mathcal K} = K + \Delta {\mathcal K} $
are cast into the term, $\Delta \Pi  (q,u)$, of (\ref{k-self}). It
is crucial to consider the latter term since it will introduce in
particular the important short range correlation effects described
by the Migdal parameters. This will be discussed in more detail
below.

With (\ref{rewrite}) the self consistent set of equations
(\ref{hatt},\ref{hatg},\ref{k-self}) is rewritten in a way that
one may start with a set of tabulated free-space scattering
amplitudes $T^{(I)}$ in terms of which self consistency is
achieved. Coupled channel effects, which are known to be important
for $\pi N$ scattering at higher energies, are included by
assigning ${\mathcal T}$, ${\mathcal G}$ and $K$ the appropriate
matrix structures. The effect of all inelastic channels can be
accounted for by a renormalization leading to a complex scattering
kernel $K$ of the $\pi N$ channel. Therefore we may assume that
the amplitudes $T^{(I)}$ in (\ref{rewrite}) already include  the
dynamics of all inelastic coupled channels. In this work we study
the consequence of the in-medium modified $\pi N$ channel
exclusively. Thus it is not necessary to make the coupled channel
structure of the $\pi N$ amplitudes explicit \cite{LuKo}.
Ultimately it would be desirable to also evaluate the in-medium
modification of the inelastic processes described for instance by
the $\pi \Delta$, $\rho N$ and $\omega N$ channels.

The scattering amplitudes can be systematically decomposed into
covariant projectors $Y^{(\pm)}_n(\bar q,q;w)$ with good angular
momentum and parity $J^P=(n+\half)^\mp$. We write
\begin{eqnarray}
T^{(I)}(\bar k,k;w) &=& \sum_{n=0}^\infty\,
Y^{(+)}_n (\bar q, q;w)\,M^{(+)}_{\,I}(\sqrt{s}\,,n)
\nonumber\\
&+& \sum_{n=0}^\infty\,
 Y^{(-)}_n (\bar q, q;w)\,M^{(-)}_{\,I}(\sqrt{s}\,,n) \,,
\label{t-vacuum}
\end{eqnarray}
where $w^2=s$ and $k= \half\,(p-q)$ and $\bar k =\half\,(\bar
p-\bar q)$. The representation (\ref{t-vacuum}) implies a
particular off-shell behavior of the scattering amplitude, which
was obtained in the course of constructing a systematic on-shell
reduction of the covariant Bethe-Salpeter equation that does not
depend on the choice of meson and baryon interpolation fields
\cite{LK}. In this work we focus on the leading $J=
{\textstyle{1\over 2}}$ and  $J= {\textstyle{3\over 2}}$ channels
with $ Y_0^{(+)}$ (s-wave), $ Y_0^{(-)}$ (p-wave) and $ Y_1^{(+)}$
(p-wave), $ Y_1^{(-)}$ (d-wave). For details on the construction
and the properties of these projectors we refer to \cite{LK,LuKo}.

For energies above threshold $\sqrt{s} > m_N+m_\pi$ the invariant
amplitudes are uniquely determined  by the scattering phase shifts
$\delta^{(l)}_J(\sqrt{s}\,)$ and inelasticity parameters
$\eta^{(l)}_J(\sqrt{s}\,)$,
\begin{eqnarray}
M^{(\pm )}_{}(\sqrt{s},J-{\textstyle{1\over2}}) &=& -i\,\frac{8
\,\pi\,s}{p^{\,2\!\,J}_{\pi N}} \, \frac{\eta^{(l)}_{J=l\pm
\frac{1}{2}}(\sqrt{s}\,)\,\Big( e^{2\,i\,\delta^{(l)}_{J=l\pm
\frac{1}{2}} (\sqrt{s}\,)}-1 \Big)} {s+m_N^2-m_\pi^2  \pm 2\,m_N
\,\sqrt{s}} \,, \nonumber\\
\frac{p_{\pi N}^2}{s} &=& \frac{1}{4}\,\Big(1-
\frac{(m_N+m_\pi)^2}{s} \Big)\,\Big(
1-\frac{(m_N-m_\pi)^2}{s}\Big) \,.
 \label{match}
\end{eqnarray}
We use $m_N= 939$ MeV and $m_\pi =139$ MeV throughout this work.
At subthreshold energies $\sqrt{s} < m_N-m_\pi$ the on-shell
amplitudes are again determined by the scattering phase shifts
once crossing symmetry is invoked. In the region $m_N-m_\pi <
\sqrt{s} < m_N+m_\pi$ an analytic continuation is required as for
instance implied by the dispersion-integral representation,
\begin{eqnarray}
\Re\,M^{(\pm)}_I (\sqrt{s}\,,n) =
D^{(\pm)}_I(\sqrt{s},n)+{\mathcal P}\int_{m_N+\,m_\pi}^\infty
\,\frac{d \,w}{\pi }\,\frac{\Im M^{(\pm )}_I(w, n)}{w -\sqrt{s} }
\,,
 \label{an-con}
\end{eqnarray}
where the functions $D^{(\pm)}_I(\sqrt{s},n)$ represent all
left-hand cut contributions \cite{Landoldt:Boernstein}. Since the
cut structure in $D_I^{(\pm)}(\sqrt{s},n)$ is largely dominated by
the s- and u-channel nucleon-exchange contributions, it is
straightforward to analytically continue the {\it on-shell}
amplitudes into the region $m_N-m_\pi <\sqrt{s}< m_N+m_\pi$ as to
arrive at approximate partial-wave amplitudes specified for all
energies. As a consequence the amplitudes would receive large and
strongly energy-dependent contributions in the interval $m_N-m_\pi
< \sqrt{s} < m_N+m_\pi $.

We point out, however, that the formulation as presented above
would lead to unphysical and misleading results for the pion self
energy unless the dependence of the partial-wave amplitudes on
$q^2 \neq m_\pi^2$ at subthreshold energies is taken into account
in some way. This is important, since the u-channel contribution
to the partial wave amplitudes shows an extremely strong
dependence on $q^2$. Typically for given $\sqrt{s}$ with
$m_N-m_\pi <\sqrt{s}< m_N+m_\pi$ and $q^2<0$ the contribution is
small and negligible but at the on-shell point $q^2 = m_\pi^2$ it
is large. Since the subthreshold amplitudes are probed by the pion
self energy only for $q^2< m_\pi^2$, it is clear that the
$q^2$-dependence must be considered. Rather than extending our
scheme for the most general off-shell structure in the scattering
amplitude we propose a simple modification of the scheme that
successfully circumvents the artifacts of a pure on-shell scheme.
It would be unclear in any case what to use for the off-shell
dependence, since such a dependence is highly scheme-dependent.

\begin{figure}[t]
\begin{center}
\includegraphics[width=14.5cm,clip=true]{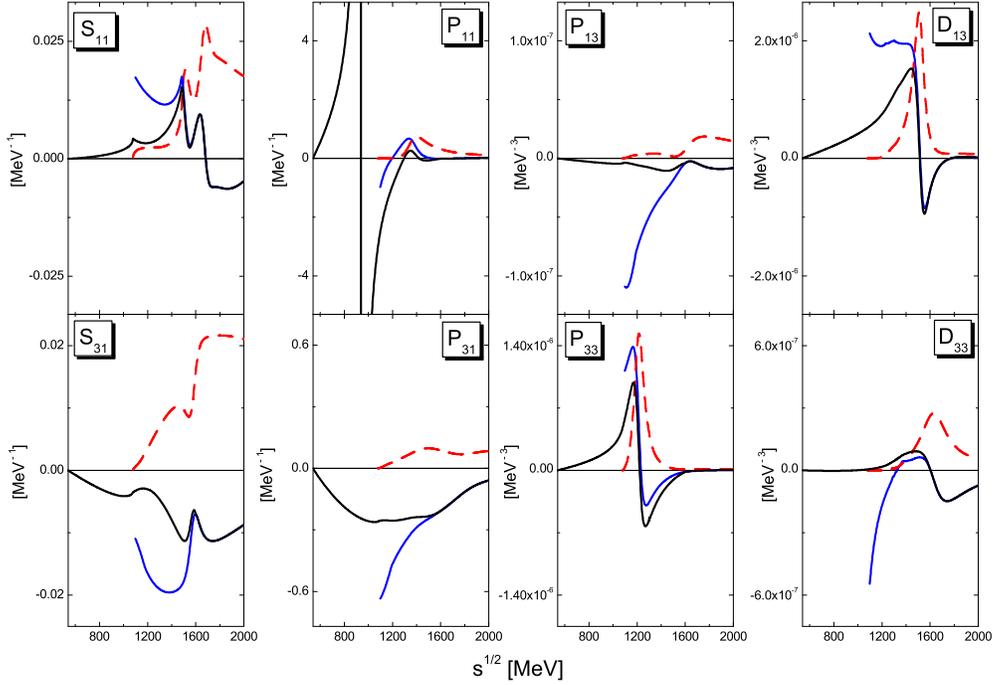}
\label{fig:1}
\end{center}
\caption{Partial wave $\pi N$ scattering amplitudes at
subthreshold energies as reconstructed from the SM01 phase shifts
\cite{Arndt}. The solid and dashed lines give the real and
imaginary parts of the $M_I^{(\pm)}(\sqrt{s}, J-1/2)$ amplitudes
of (\ref{match}). The additional solid lines extending down to
subthreshold energies represent the amplitudes $\bar
M_I^{(\pm)}(\sqrt{s}, J-1/2)$ that represent $T^{(I)}_s (\bar q,
q; w)$ of (\ref{def-decom}) and are free of u-channel cuts.}
\label{fig:pion-amplitudes}
\end{figure}

The idea the proposed scheme is based on, exploits crossing
symmetry as a tool that determines the troublesome off-shell part
of the scattering amplitude. We seek a representation of the
free-space scattering amplitude of the form
\begin{eqnarray}
&&T^{(\frac{1}{2})}_s (\bar q, q; w)=-{\textstyle{1\over
3}}\,T^{(\frac{1}{2})}_u (-q,-\bar q;w-q-\bar q)
+{\textstyle{4\over 3}}\,T^{(\frac{3}{2})}_u
(-q,-\bar q;w-q-\bar q)\,, \nonumber\\
&&T^{(\frac{3}{2})}_s (\bar q, q; w)=+{\textstyle{2\over
3}}\,T^{(\frac{1}{2})}_u (-q,-\bar q;w-q-\bar q)
+{\textstyle{1\over 3}}\,T^{(\frac{3}{2})}_u (-q,-\bar q;w-q-\bar
q)\,, \nonumber\\
&& T^{(I)}(\bar k, k;w) = T^{(I)}_s (\bar q, q;w)+T^{(I)}_u (\bar
q, q;w)  \,, \label{def-decom}
\end{eqnarray}
that manifests the desired crossing symmetry of the amplitudes
explicitly but masks the constraints set by unitarity. The
decomposition (\ref{def-decom}) is not unique per se but can be
tailored to our application. In particular we insist that $T_s$
contains only s-channel unitarity cuts but shows no cuts in the
u-channel. Similarly we insist that $T_u$ is real for $\sqrt{s}>
m_N+ m_\pi$. In general there will be contributions present in $T$
which lead to both an s-channel and u-channel cut. Such
contributions must be put into a remainder term not displayed in
(\ref{def-decom}), if we wish to keep perfect analytic properties
of the amplitudes $T_{s}$ and $T_u$. We point out however, that
such contributions are zero identically, if one considered elastic
scattering processes only. Therefore we expect the degree of
analyticity violation, when demanding $T_s$ to represent the
complete strength of the s-channel unitarity cut, to be reasonably
small if energies are not excessively large. Within a given
application window there is still the freedom how to distribute a
polynomial background contribution. Here it is advantageous to
insist on $T_s = T$ beyond a certain scale $\sqrt{s}> \Lambda$.
Consistency then requires that for $\sqrt{s} <
(m_N^2-m_\pi^2)/\Lambda$ one should obtain $T_s = 0 $ or
equivalently $T = T_u$. The scattering amplitude $T_s$ is
decomposed into partial wave amplitudes $\bar
M^{(\pm)}_{I}(\sqrt{s},n)$. In practise we use the representation
(\ref{an-con}) with the function $D^{(\pm)}_I(\sqrt{s},n)$
replaced by a second order polynomial in $\sqrt{s}$.  The
coefficients of the latter  are determined to guarantee first,
that the real parts of $\bar M^{(\pm)}_{I}(\sqrt{s},n)$ vanish
identically at $\sqrt{s}= (m_N^2-m_\pi^2)/\Lambda$ and second, is
such that a smooth matching to the real part of the partial wave
amplitudes, $M^{(\pm)}_{I}(\sqrt{s},n)$, at $ \sqrt{s}= \Lambda $
as given in (\ref{match}) is obtained. Of course in the $P_{11}$
channel the nucleon pole contribution is incorporated in addition.
The pion-nucleon coupling constant, $g_{\pi NN } = 13.34$, of
\cite{Ericson} is used. With $\Lambda = 1600$ MeV  a remarkably
smooth matching is achieved for all considered channels as
described above. The deviation of $\bar M^{(\pm)}_{I}(\sqrt{s},n
)$ from $M^{(\pm)}_{I}(\sqrt{s},n )$, both shown in Fig.
\ref{fig:1}, is a measure for the importance of u-channel cuts in
a given partial wave.

The idea is to use the amplitudes $T_s$ in (\ref{def-decom})
properly decomposed into partial waves via (\ref{t-vacuum}). The
effect of $T_u$ is incorporated into the pion self energy by
adding to (\ref{k-self}) the term induced by $T_s$ but with $q \to
-q$ replaced. As we apply the decomposition (\ref{def-decom}) for
a many-body evaluation of the pion self energy, the parameter
$\Lambda$ plays the role of a scale that determines at what energy
the manifestly crossing-symmetric scheme recovers two-body
unitarity exactly. For energies close to the pion-nucleon
threshold a unitarization is not really required for a
quantitative description of the scattering amplitude. The
amplitudes are largely dominated by the s- and u-channel baryon
exchange contributions. In contrast, at higher energies the
unitarity constraint becomes more and more important rendering any
constraint from crossing symmetry rather implicit.

The solution of the in-medium Bethe-Salpeter equation is
considerably complicated by the presence of further tensor
structures $P_{[ij]}$ and $Q_{[ij]}$ not required in the vacuum.
For a complete collection of the in-medium projectors together
with their induced in-medium loop functions, $\Delta
J^{(p,q)}_{[ij]}(w , u)$, representing the object $\Delta
{\mathcal G}$ in (\ref{rewrite}), see \cite{LuKo}. In this work
the loop functions are evaluated by their dispersion-integral
representation in terms of their imaginary parts. A subtraction
constant is determined by insisting that the scattering amplitudes
do not renormalize the nucleon mass
\begin{eqnarray}
\Delta J^{(p,q)}_{[ij]}(w,u)\Big|_{w^2=m_N^2} = 0 \,\;\to \quad
{\mathcal T}_s^{(I)}(\bar q,q;w,u)\Big|_{w^2=m_N^2}=T_s^{(I)}(\bar
q, q,w)\,, \label{def-sub-g}
\end{eqnarray}
for $ w \cdot u > 0 $. Such a condition is well justified in the
present scheme since neither correlations nor binding effects are
incorporated. That is outside the scope of this work. Furthermore
we drop a small contribution of intermediate hole states, not
consistently treated here in any case, but part of the
relativistic scattering equation (\ref{hatt}). This leads to an
in-medium scattering amplitude, whose imaginary part,
\begin{eqnarray}
\Im \,{\mathcal T}_s(\bar q,q;w,u) =0 \quad  {\rm if}\; \quad w
\cdot u < \sqrt{m_N^2+k_F^2} \;, \label{t-pauli}
\end{eqnarray}
vanishes below the chemical potential as expected from the Pauli
principle. To be precise the property (\ref{t-pauli}) requires
also the consideration of the Pauli blocking effect in the
s-channel nucleon exchange contribution, i.e. ${\mathcal K} \neq
K$.

We turn to the effects of short range correlation \cite{Migdal}.
Here we follow the recent work \cite{Lutz:Migdal} and apply
covariant expressions parameterized in terms of the Migdal
parameters $g'_{11}, g'_{12}$ and $g'_{22}$. The delta-hole term
is folded by an isobar spectral function obtained from the
appropriate in-medium scattering amplitude assuming the averaged
value  $\vec w=200$ MeV. We then approximate the isobar spectral
function to be a function of $w_0^2-\vec w^2$. Thus, the value for
$g_{22}'$ used in this work is introduced with respect to the
in-medium pion-isobar coupling constant. The latter will show a
sizeable in-medium reduction in our scheme. The expression for the
pion self energy as given in \cite{Lutz:Migdal} properly
subtracted by its leading order contribution, defined in the limit
$g'_{ij}\to 0$, is identified with $\Delta \Pi$ in (\ref{k-self}).
The real part of this contribution is evaluated in terms of a
dispersion-integral representation with a subtraction constant
fixed at $\omega^2-\vec q\,^2 = m^2_\pi $ matched to the real part
of the original $\Delta \Pi$ contribution. Since the nucleon-hole
contribution as derived in terms of the in-medium scattering
amplitude includes an appreciable momentum dependent
renormalization of the pion-nucleon coupling constant, a
corresponding renormalization is applied in $\Delta \Pi$. This
effect reduces the strength of the nucleon-hole contribution. The
values we quote here for Migdal's $g'_{11}$ and $g'_{12}$
parameters are defined with respect to the free-space pion-nucleon
but in-medium pion-isobar coupling constants \cite{Lutz:Migdal}.

We reconstruct the real part of the pion self energy, $\Re \Pi
(q,u)$, in terms of its imaginary part, $\Im \Pi (q,u)$. From now
on we work in the rest frame of nuclear matter with $u_\mu=(1,0)$
for convenience and write $\Pi (q,u) = \Pi (\omega, \vec q\,)$. A
subtracted dispersion-integral representation is imposed
\begin{eqnarray}
\Pi  (\omega , \vec q\,) = c (\vec q\,)+ \int_{0}^\lambda
\,\frac{d \,\bar \omega^2}{\pi }\, \frac{\Im \,\Pi (\bar \omega ,
\vec q\,)}{\bar \omega^2 -\omega^2 -i\,\epsilon}
\,,\label{pi-disp}
\end{eqnarray}
cut off by $\lambda =1.2$ GeV. The reflection property $\Pi
(\omega, \vec q\,)=\Pi  (- \omega, \vec q\,)$ holds for isospin
symmetric nuclear matter as a direct consequence of our manifestly
crossing-symmetric scheme. The  subtraction constant $c (\vec
q\,)$ is fixed such that the real part of the self energy
reproduces the corresponding value for the contribution of the
amplitude defined by (\ref{k-self}) at the point
$$
\omega = {\rm Max} [\sqrt{\Lambda^2+(|\vec
q\,|+k_F)^2}-\sqrt{m^2_N+k_F^2}, \sqrt{\Lambda^2+|\vec q\,|^2}-m_N
]\,.
$$
This condition guarantees that the real part of the in-medium
amplitude is probed only for $w_0^2-\vec w^2 \geq \Lambda^2$,
where the effective amplitude $T^{(I)}_s (\bar q, q;w)$ represents
the exact scattering amplitude (see (def-decom)). We assure that
the sum rule
\begin{eqnarray}
\int_{-\infty}^{+\infty} \frac{d \omega}{\pi}\,|\omega | \,\Im \,D
(\omega ,\vec q \,) = -1 \,,\label{sum-rule}
\end{eqnarray}
holds accurately within 1 $\%$ in our scheme.

The self consistent set of equations (\ref{k-self},\ref{hatt}) is
solved numerically by iteration where we start with the evaluation
of the in-medium self energy. Convergence is typically found after
5 to 6 iterations. The energies and the three-momenta are
restricted by $|\omega| < 1.2$ GeV, $|\vec q\,| < 0.8$ GeV and
$w_0 < 2$ GeV, $|\vec w| < 1.2$ GeV. The free-space partial wave
amplitudes, $\bar M^{(\pm)}_{I}(\sqrt{s},n)$, are put to zero for
$\sqrt{s}>2$ GeV and $\sqrt{s}<0.54$ GeV.

\section{Results}

We give a presentation and discussion of our results for the
in-medium modification of the pion and the
$J=\frac{1}{2},\frac{3}{2}$ nucleon and isobar resonance
properties. The resonance propagator can be identified with the
appropriate pion-nucleon scattering amplitude of a given partial
wave. In the self consistent scheme of section 2 the in-medium
scattering process is intimately related to the pion spectral
function. According to (\ref{k-self}) once the self consistent
in-medium scattering process is established the pion self energy
follows by averaging the in-medium scattering amplitudes
(\ref{k-self}) over the Fermi distribution. Therefore the
pertinent structures in the in-medium amplitudes already tell the
characteristic features expected in the pion spectral function.

Since the present day literature does not offer a unique set of
Migdal parameters $g'_{11}, g'_{22}$ and $g'_{12}$ we studied
first the sensitivity of our approach to different choices
thereof. We use two sets of parameters in this work. Set I) is
assuming $g'_{11}=g'_{12}=g'_{22}=0.8$. Set II)
is given by the values $g_{11}' = 0.585\,, g_{12}' = 0.15$ together with $g_{22}' = 0.60$.
We remind the reader that we use here a convention in which
$g'_{11}$ is defined with respect to the free-space pion-nucleon coupling constant
whereas $g'_{22}$ is defined with respect to the in-medium pion-nucleon-isobar coupling
constant.

\begin{figure}[t]
\begin{center}
\includegraphics[width=14cm,clip=true]{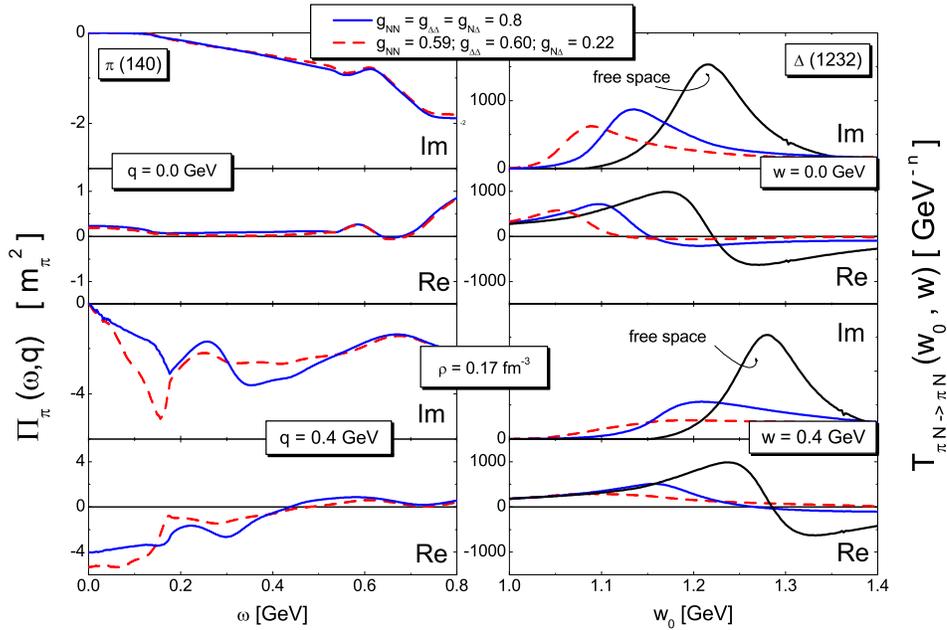}
\end{center}
\caption{In-medium pion self energy (left hand panels) and $\pi N$
scattering amplitude in the $P_{33}$ channel (right hand panels)
at nuclear saturation density. The solid lines show the results
obtained with set I) of Migdal parameters, the dashed lines those
for set II) as specified in the text. The upper two (lower two)
right hand panels describe isobar propagation in nuclear matter
with total isobar three momentum $|\vec w| =0.0$ GeV ($|\vec w|=
0.4 $ GeV), compared to free space results. }\label{fig:2}
\end{figure}

With the nuclear density set to $\rho =0.17$ fm$^{-3}$ Fig.
\ref{fig:2} shows the resulting pion self energy together with the
real and imaginary parts of the in-medium $\pi N$ scattering
amplitude in the $P_{33}$ channel. As illustrated by the figure
the effect of different choices for the Migdal parameters affects
the pion self energy mostly at not too large energies and
intermediate momenta. As illustrated by the right hand panels the
effects of such differences are quite important for the properties
of the $\Delta(1232)$-isobar in nuclear matter. Only set I) leads
to acceptable properties of the isobar though we may still
overestimate the broadening of its width in nuclear matter for
high velocities. Note that the depletion of the isobar peak in
Fig. \ref{fig:2} is a combined effect of reduced pion-isobar
coupling constant and increased decay width. Thus the $g'_{22}$
parameter as defined more conventionally with respect to the
free-space pion-nucleon-isobar coupling constant is in fact
smaller than 0.8 about 0.6 only. For isobar propagation with
non-zero momenta set II) leads to a substantial broadening
unlikely to be consistent with photo-absorbtion data of the
nucleus \cite{photo-absorption}. We confirm the results of
\cite{Oset:Salcedo} that the splitting of transverse (Q-space) and
longitudinal (P-space) modes is small. In our formulation such
effects reflect the in-medium mixing of partial wave amplitudes.
In none of the channels we observed significant effects thereof.
The attractive mass shift for the isobar of about 80 MeV is in
qualitative agreement with previous microscopic calculations
\cite{Oset:Salcedo}. Referring to the argument put forward in
\cite{Oset:Weise,Oset:Salcedo} the apparent repulsive mass shift
as extracted from photo absorption data \cite{Thies} is a combined
effect of an attractive isobar self energy and short range
correlation effects.

The results differ from conclusions of previous works
\cite{Xia:Siemens:Soyeur,Korpa:Malfliet,Knoll} that also
incorporated self consistency but did not observe such a
significant broadening of the isobar. This difference is a direct
consequence of the soft form factor in the pion-isobar vertex used
in \cite{Xia:Siemens:Soyeur,Korpa:Malfliet,Knoll}. In our scheme
there is not much place for such phenomenology, since the applied
phase shifts entail already  form factor effects that one may want
to use modeling the pion-nucleon interaction. We emphasize our
goal to develop a {\it microscopic} understanding of pion
propagation in nuclear matter. First we continue to present and
discuss results solely based on the $\pi N$ phase shifts and a set
of Migdal parameters only. Nevertheless, we will return to this
issue and also present results that follow upon incorporating a
phenomenological form factor into our scheme.

\subsection{Propagation without form factors}

In the form factor-free scheme the isobar width can be protected
against excessive increase to some extent by an appropriate choice
of Migdal parameters. Large $g'_{12}$ and $g_{22}'$ parameters
together with a reasonably large $g_{11}'\sim 0.6-0.8$ lead to a
suppression of fast soft modes that are responsible for the
broadening of the isobar, i.e. the low-energy tail of the pion
spectral function at momenta $|\vec q\,| \sim 400-500$ MeV is
suppressed. We are aware that this conclusion is possibly
inconsistent with the conclusion of the recent work \cite{Suzuki}.
This issue certainly deserves more detailed investigations. Vertex
correction diagrams that mask the pion spectral function as probed
in the isobar self energy but not considered in the present scheme
yet may lead to a similar suppression of fast soft modes. Thus we
would not exclude a small value of $g_{12}'$ once such effects are
incorporated.

A further interesting question is the amount of softening found in the pion self energy.
Typically the traditional nucleon and isobar-hole model leads to a minimum of the function
\begin{eqnarray}
S(\vec q\,)= \vec q\,^2 + m_\pi^2+\Pi (0, \vec q)\,,
\label{}
\end{eqnarray}
at some intermediate momentum. Such a minimum helps to explain for
instance unnatural parity states of finite nuclei \cite{Migdal}.
To be specific the covariant nucleon and delta-hole model proposed
in \cite{Lutz:Migdal} shows a minimum at $|\vec q\,| \simeq 365$
MeV and nuclear saturation density using the parameter set II). In
the present self consistent scheme we do not find a minimum of
$S(\vec q\,)$ for either of the two choices of Migdal parameters
studied here. The size of the function at a given momentum, say
$|\vec q\,|= 300$ MeV, strongly depends  on the choice of Migdal
parameters, with $S\simeq 3.4\,m_\pi^2 $ and $S \simeq
2.4\,m_\pi^2$ for set I) and II) respectively. By lowering
$g'_{11}$ and $g'_{12}$ down to unrealistic values it is possible
to produce a minimum of $S(\vec q)$ in our present scheme, however
at the prize that the isobar resonance is dissolved almost
completely.

\begin{figure}[t]
\begin{center}
\includegraphics[width=14cm,clip=true]{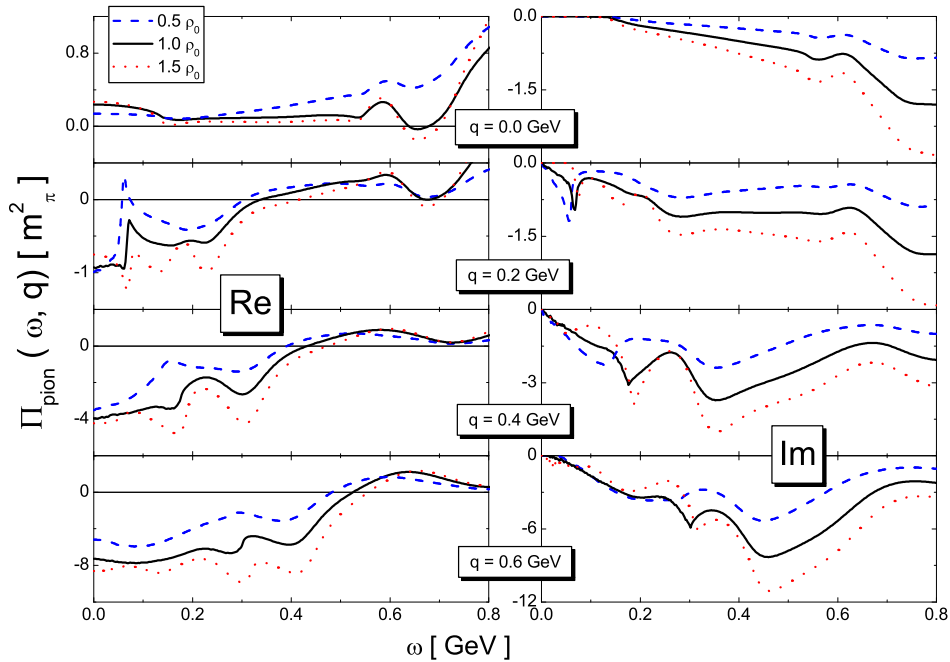}
\end{center}
\caption{Real (left hand panel) and imaginary (right hand panel)
part of the pion self energy at $\rho = 0.5 \,\rho_0$, $\rho_0$
and $1.5 \,\rho_0$ with $\rho_0= 0.17$ fm$^{-3}$. The results were
obtained with parameter set I).} \label{fig:pions-05-10-15}
\end{figure}

At small pion momenta $|\vec q\,| < 40 $ MeV the pion self energy
may be approximated by
\begin{eqnarray}
\Pi (\omega,\vec q\,)\Big|_{\omega^2 = \Re\,[\alpha\,m_\pi^2+\beta\,\vec q\,^2\,]}
\simeq (\alpha-1)\,m_\pi^2 +(\beta-1)\,\vec q\,^2 \,,
\label{pi-param}
\end{eqnarray}
the values of the parameters known to some extent from pionic atom
data \cite{Ericson:Weise,Nieves:Oset:Recio}. For set I) and II) we
obtain $(\alpha , \beta)\simeq (1.09-i\,0.05, 0.56 -i\,0.25)$ and
$(\alpha , \beta) \simeq (1.06-i\,0.07, 0.74 -i\,0.44)$. We do not
find any significant effect from a possible wave function
renormalization. Comparing (\ref{pi-param}) with the results of
\cite{Nieves:Oset:Recio} the s-wave absorption strength
parameterized by $\Im \,\alpha $ comes out about a factor two too
small. This reflects the fact that not all absorption channels are
included in our work. The value for $\Re\,\alpha $ is reasonably
close to phenomenological values. The p-wave absorption strength
related to $\Im \,\beta$ is overestimated somewhat
\cite{Ericson:Weise}. This may be linked to a possibly too large
in-medium isobar width obtained in our present scheme. The
parameter $\Re \,\beta$ of set I) is however close to the
empirical value \cite{Ericson:Weise}. Given the fact that our
scheme is parameter free except for the choices of the Migdal
parameters, these results are encouraging. However, it should be
emphasized that one expects some non-linear dependence of $\alpha
$ and $\beta $ on the density. For instance at $0.5 \,\rho_0$ we
obtain $(\alpha , \beta )\simeq (1.09-i\,0.02, 0.46-i\,0.11)$ for
set I). To further illustrate the amount of non-linear behavior in
the pion self energy we include Fig. \ref{fig:pions-05-10-15}
which displays the results for the pion self energy for the
nuclear densities $0.5 \,\rho_0$, $1.0\,\rho_0$ and $1.5\,\rho_0$.
A striking non-linear behavior is typically seen at small pion
energies.

\begin{figure}[t]
\begin{center}
\includegraphics[width=14cm,clip=true]{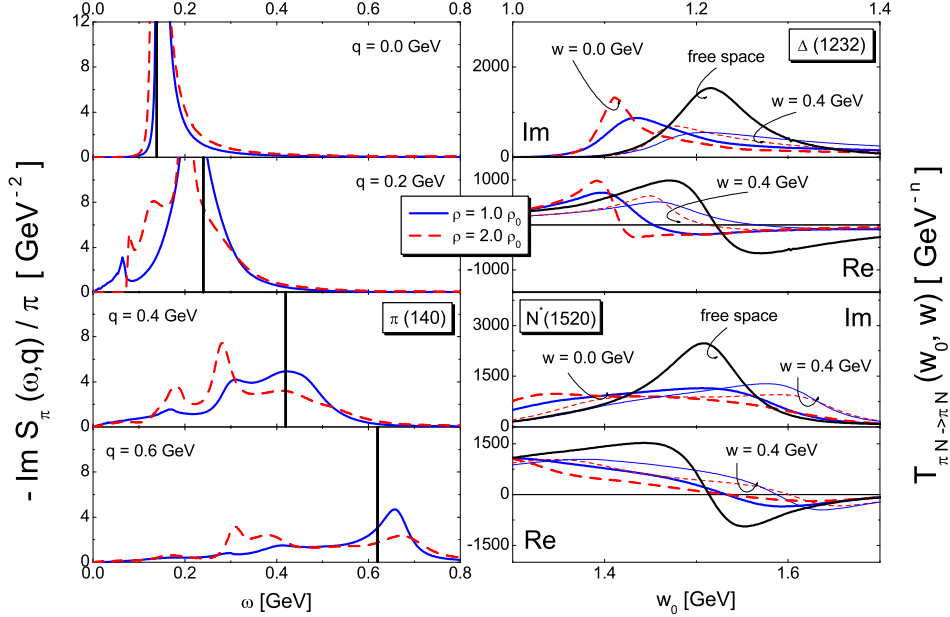}
\end{center}
\caption{Pion spectral function (left hand panels) and
$J=\frac{3}{2}$ nucleon resonance propagators (right hand panels)
at $\rho_0= 0.17$ fm$^{-3}$ (solid lines) and $2\,\rho_0$ (dashed
lines) as functions of the pion, $\omega$, and resonance, $w_0$,
energy respectively. The vertical lines in the left hand panels
show the energy of a pion in free space at given momentum $|\vec
q\,|$. The upper (lower) two right panels show the in-medium $\pi
N$ scattering amplitude in the $P_{33}$ ($D_{13}$) channel.
Results are shown for a resonance at rest with $|\vec w|=0$ (thick
lines) and for three-momentum $|\vec w| = 0.4$ GeV (thin lines)
with longitudinal polarization. } \label{fig:pion-nstar-spec}
\end{figure}

In Fig. \ref{fig:pion-nstar-spec} our results for the pion
spectral function and the $J=\frac{3}{2}$ isobar $\Delta(1232)$
and $N(1520)$ resonances are shown for set I) of Migdal parameters
at two different nuclear densities, $\rho_0= 0.17$ fm$^{-3}$ and
$2\,\rho_0$. Of course the results at $2 \,\rho_0$ should be
considered cautiously because nuclear binding and correlation
effects were not yet fully included  in the present scheme. The
pion spectral function clearly exhibits  the three well known
modes, zero-sound, isobar-hole and pion branch with weighting
factors strongly dependent on the pion momentum. We wish to make
two points here. First, in contrast to the standard delta-hole
model for the pion \cite{Oset:Weise}, density we do not observe a
significant strength of soft pion modes at nuclear saturation in
our work. This is due to self consistency. Using somewhat smaller
values $g_{ij}'=0.6$ does not change our conclusion qualitatively.
The resulting spectral function is quite similar to the one shown
in Fig. \ref{fig:pion-nstar-spec} only that  for instance at
$|\vec q\,|= 400$ MeV about 25 $\%$ of the strength sitting in the
pion branch is moved up to the delta-hole branch and to a lesser
degree into the zero-sound branch. Our finding confirms the
results of previous self consistent approaches to pions in nuclear
matter \cite{Korpa:Malfliet,Xia:Siemens:Soyeur,Knoll}
qualitatively and should have important consequences in various
applications of the pion spectral function to hadron properties in
nuclear matter. Second, the significant broadening and repulsive
shift of the main mode at large momenta is obviously a result of
the inclusion of s- p- and d-waves in our scheme. Such effects are
lacking in a model that incorporates nucleon- and delta-hole terms
only.

We turn to the resonance properties. Fig.
\ref{fig:pion-nstar-spec} shows the narrowing of the isobar
resonance with increasing density due to the growing importance of
Pauli blocking. A similar  but much less pronounced effect is
observed for the $N(1520)$ resonance. The significant broadening
of the latter resonance at nuclear saturation density is in
qualitative agreement with previous phenomenological works
\cite{Kondratyuk,Alberico,Chen,Boffi} and may help to arrive at a
microscopic understanding of photo absorption data in the second
resonance region expected to be dominated by the contribution of
the $N(1520)$ resonance \cite{Krusche,Effenberger,Lehr}. The fact
that at nuclear saturation density we do not obtain any large
in-medium effect from the $\pi N$ channel on the remaining
resonances, $N(1535)$, $N(1650)$, $\Delta(1620)$, $\Delta (1600)$
and $\Delta (1700)$ except for the $N(1440)$, for which we find a
significantly increased width, is interesting and deserves further
studies. The $D_{13}$ and $P_{11}$ resonances $N(1520)$ and
$N(1440)$ appear particularly sensitive to the in-medium dressing
of the $\pi N$ channel due to their d-wave and p-wave phase space
behavior. Clearly the effect of possible in-medium modifications
of inelastic channels like $\pi \Delta, \rho N, \omega N $ on the
resonance properties asks for further detailed studies
\cite{Peters}.

\subsection{Propagation with form factors}

\begin{figure}[t]
\begin{center}
\includegraphics[width=14cm,clip=true]{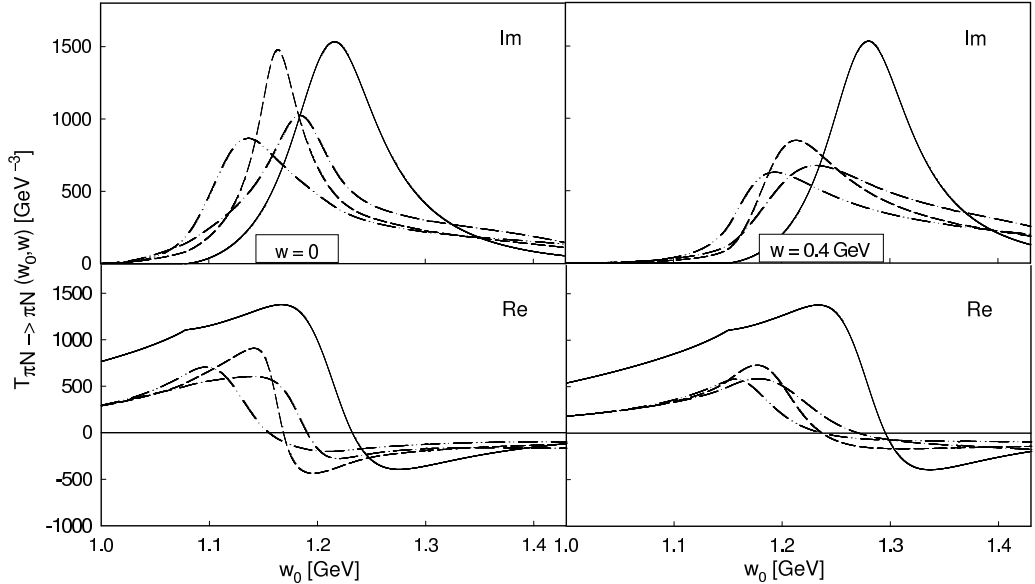}
\end{center}
\caption{In-medium $\pi N$ scattering amplitude in the $P_{33}$
channel at nuclear saturation density, compared to the vacuum
amplitude (solid line). The values of Migdal parameters are:
$g'_{11}=g'_{12}=g'_{22}=0.8$. The dashed line corresponds to
$\Lambda_{\pi NN}=0.5\;$GeV and $\Lambda_{\pi N\Delta}\rightarrow
\infty$, the dash-dot line to $\Lambda_{\pi NN}=\Lambda_{\pi
N\Delta}=0.5\;$GeV, the dash-dot-dot line to the calculation
without form factors. The left two (right two) panels describe
isobar propagation in nuclear matter with total isobar three
momentum $|\vec w| =0$ GeV ($|\vec w|= 0.4 $ GeV). }
\label{delta1}
\end{figure}

Until now we did not include any phenomenological form factors
into the computation, motivated by the fact that most of such
effects are already taken care of by using the physical scattering
amplitudes. However, a certain form of pion-nucleon-nucleon and
pion-nucleon-isobar form factors commonly used in calculations of
pion, nucleon and isobar self energies based on the relevant
3-point functions can also be introduced in the present scheme.
The condition is that the solution of the Bethe-Salpeter equation
in vacuum is not affected for the on-shell scattering amplitudes.
Indeed, the use of scattering phase shifts incorporates the form
factors necessary to describe the pion-nucleon interaction, but
only for on-shell particles. Off-shell-pion effects may allow the
inclusion of phenomenological form factors of the form,
\begin{equation}
F_{\pi NX}(q^2)=\exp \left[ -(q^2-m^2_\pi)^2 /\Lambda^4_{\pi NX}\right],
\end{equation}
which suppress off-shell-pion contributions from relevant channels
(with $X$ being $N$ or $\Delta$) when calculating the pion self
energy. It should be emphasized, however, that the
phenomenological need for such a form factor may stem from medium
modifications of the interaction vertices rather than a strong
off-shell dependence of the vertices in free-space. We do not
modify the loop-integral calculation, since for consistency that
would require considering also the influence of the Migdal
parameters on the pion-nucleon-nucleon and pion-nucleon-isobar
vertices in those loop integrals. This is not considered in this
work due to serious complications.

\begin{figure}[t]
\begin{center}
\includegraphics[width=14cm,clip=true]{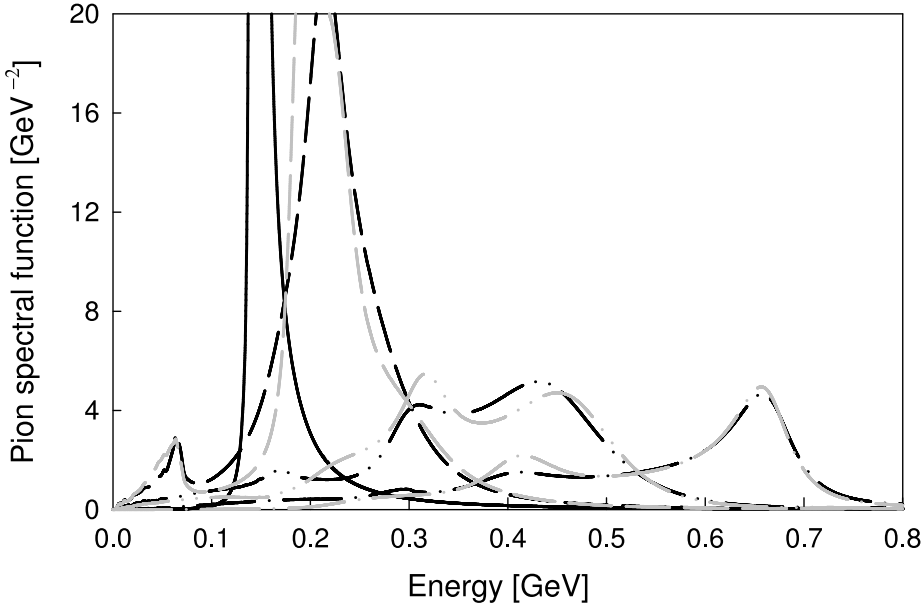}
\end{center}
\caption{In-medium pion spectral function at nuclear saturation
density, for different values of pion momentum: 0 (solid line),
200 MeV (long dash), 400 MeV (short dash), 600 MeV (dot-dash
line). The Migdal parameters are $g'_{11}=g'_{12}=g'_{22}=0.8$.
The black line corresponds to calculation without form factors,
while the gray lines are obtained with $\Lambda_{\pi
NN}=0.5\;$GeV, $\Lambda_{\pi N\Delta}\rightarrow\infty$. For zero
momentum the change is marginal and not shown. } \label{pion2}
\end{figure}

In Fig.~\ref{delta1} we show the effect of the form factors on the
$\pi N$ scattering amplitude in the $P_{33}$ channel at nuclear
saturation density (other channels are affected marginally).
Whereas the form-factor influence is quite pronounced at momentum
$|\vec w|=0$ it is of minor importance at $|\vec w|=0.4\;$GeV.  As
expected, the $\pi NN$ form factor reduces the strength of the
pion spectral function at low energy (less than $m_\pi$), thus
suppressing the isobar decay amplitude and decreasing its width.
The influence of the $\pi N\Delta$ form factor is less pronounced,
since the delta-hole branch of the pion spectral function is not
so far off-shell. In the present approach it somewhat broadens the
isobar and also makes its mass shift less pronounced.

\begin{figure}[t]
\begin{center}
\includegraphics[width=14cm,clip=true]{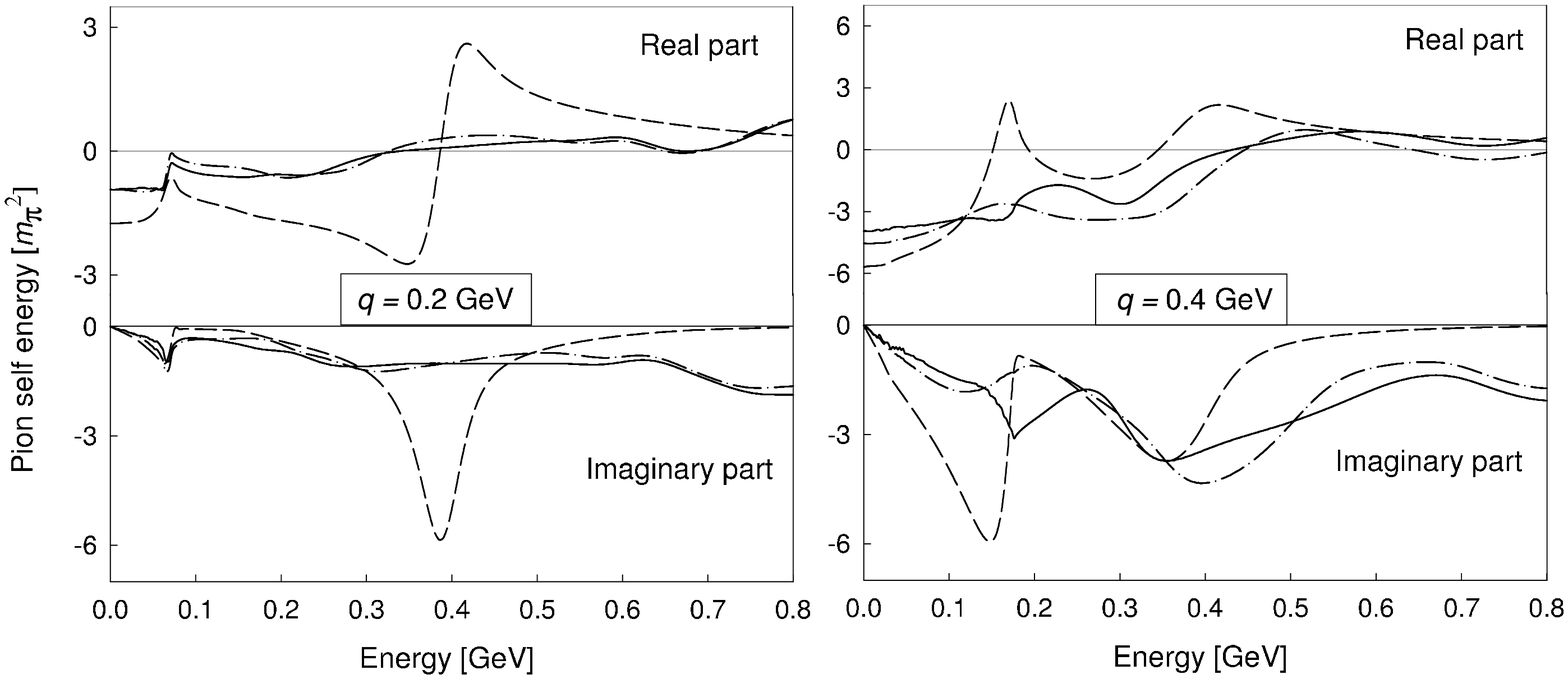}
\end{center}
\caption{In-medium pion self energy at nuclear saturation density,
for $|\vec q|=0.2\;$GeV and $|\vec q|=0.4\;$GeV pion momentum. The
full line corresponds to a calculation without form factors with
$g'_{11}=g'_{12}=g'_{22}=0.8$. The dash-dot line shows the results
for $\Lambda_{\pi NN}=\Lambda_{\pi N\Delta}=0.5\;$GeV and
unchanged $g'$ values. For comparison we show the results (dashed
line) of a non-relativistic computation from
ref.~\cite{Korpa:Malfliet}. } \label{pion-self}
\end{figure}

The pion spectral function is not dramatically changed  by the
form factors, whose effect is illustrated in Fig.~\ref{pion2}. The
figure shows the effect of the $\pi NN$ form factor since it
produces the most pronounced effect. In general we get slightly
more softening of the main pion mode and a suppression of the
very-low-energy part. The pion-nucleon-isobar form factor does not
bring in noticeable softening, although it makes the medium-energy
part of the spectral function more filled in, without suppressing
the small-energy region.

In Fig.~\ref{pion-self} we show the pion self energy at nuclear
saturation density, for momenta $|\vec q|=0.2\;$GeV and  $|\vec
q|=0.4\;$GeV. The most prominent feature of the self energy is the
strong suppression of the delta-hole contribution for momenta
around 0.2 GeV, as well as the contributions of higher-mass
resonances which leave the imaginary part non-vanishing even at
energies up to 1 GeV.

These results show that further investigations of the effects of
vertex modifications in the medium are required, which respect the
features intrinsic to our present scheme.

\section{Summary}

In this work we evaluated the pion self energy in nuclear matter
in a self consistent and covariant manner. A novel framework based
on the $\pi N$ phase shifts, as measured in free space, together
with a set of Migdal parameters was developed. Important
constraints of crossing symmetry and unitarity were incorporated
approximatively. Using reasonable values for the Migdal parameters
we found that the nucleon resonances $N(1535)$ and $N(1650)$ are
basically unaffected by the nuclear environment. Contrasted
results were obtained for the p-wave $N(1440)$ and d-wave
$N(1520)$ resonances for which we predict considerable broadening
already at nuclear saturation density.

Our result for the isobar resonance are not satisfactory at this
stage, due to a significant overestimate of its in-medium decay
width. Improved results were obtained by incorporating a soft
phenomenological form factor into the $\pi NN$ vertex. Whereas the
properties of slow isobars in nuclear matter are changed
significantly, a soft form factor has rather moderate effects on
the pion spectral function. These results show that further
detailed investigations of the effects of vertex modifications in
the medium are required to arrive at a fully microscopic
understanding of the properties of isobars in nuclear matter.

{\bfseries{Acknowledgments}}

This research was supported in part by the Hungarian Research
Foundation (OTKA) grant T030855. M.F.M. L. acknowledges useful
discussions with E.E. Kolomeitsev,  E.E. Saperstein and D.
Voskresensky. C.L.K would like to thank the NWO (Netherlands) for
providing a visitors stipend and the K.V.I. (Groningen) for the
kind hospitality.

\end{document}